\documentclass[twocolumn]{aastex62}
\usepackage[section]{placeins}
\usepackage{longtable}
\usepackage{hyperref}
\usepackage{float}
\usepackage{url}
\graphicspath{{./}{figures/}}

\shorttitle{SN\,2018lfe}
\shortauthors{Yin et al.}

\begin{document}

\title{Optical Observations and Modeling of the Superluminous Supernova 2018lfe}

\correspondingauthor{Yao Yin}
\email{yyin@college.harvard.edu}

\author[0000-0001-6395-6702]{Yao Yin}
\affil{Center for Astrophysics \textbar{} Harvard \& Smithsonian, 60 Garden Street, Cambridge, MA 02138, USA}

\author[0000-0001-6395-6702]{Sebastian Gomez}
\affil{Space Telescope Science Institute, 3700 San Martin Drive, Baltimore, MD 21218, USA}
\affil{Center for Astrophysics \textbar{} Harvard \& Smithsonian, 60 Garden Street, Cambridge, MA 02138, USA}

\author[0000-0002-9392-9681]{Edo Berger}
\affil{Center for Astrophysics \textbar{} Harvard \& Smithsonian, 60 Garden Street, Cambridge, MA 02138, USA}

\author[0000-0002-0832-2974]{Griffin Hosseinzadeh}
\affil{Steward Observatory, University of Arizona, 933 N. Cherry Ave., Tucson, AZ 85721, USA}
\affil{Center for Astrophysics \textbar{} Harvard \& Smithsonian, 60 Garden Street, Cambridge, MA 02138, USA}

\author[0000-0002-2555-3192]{Matt Nicholl}
\affil{Birmingham Institute for Gravitational Wave Astronomy and School of Physics and Astronomy, University of Birmingham, Birmingham B15 2TT, UK}

\author[0000-0003-0526-2248]{Peter K. Blanchard}
\affil{Center for Interdisciplinary Exploration and Research in Astrophysics and Department of Physics and Astronomy, \\Northwestern University, 1800 Sherman Ave, Evanston, IL 60201, USA}

\begin{abstract}
We present optical imaging and spectroscopy of SN\,2018lfe, which we classify as a Type I superluminous supernova (SLSN-I) at a redshift of $z = 0.3501$ with a peak absolute magnitude of $M_r\approx -22.1$ mag, one of the brightest SLSNe discovered. SN\,2018lfe was identified for follow-up using our FLEET machine learning pipeline. Both the light curve and the spectra of SN\,2018lfe are consistent with the broad population of SLSNe. We fit the light curve with a magnetar central engine model and find an ejecta mass of $M_{\rm ej}\approx 3.8$ M$_\odot$, a magnetar spin period of $P\approx 2.9$ ms and a magnetic field strength of $B_{\perp}\approx 2.8\times 10^{14}$ G.  The magnetic field strength is near the top of the distribution for SLSNe, while the spin period and ejecta mass are near the median values of the distribution for SLSNe. From late-time imaging and spectroscopy we find that the host galaxy of SN\,2018lfe has an absolute magnitude of $M_r\approx -17.85$, ($L_B \approx 0.029$ $L^*$), and an inferred metallicity of $Z\approx 0.3$ Z$_\odot$, and star formation rate of $\approx 0.8$ M$_\odot$ yr$^{-1}$.
\end{abstract}


\section{Introduction} 
\label{sec:intro}
Superluminous supernovae (SLSNe) are a rare class of core-collapse supernovae (SNe) that can exceed the luminosities of normal SNe by two orders of magnitude \citep{Chomiuk11, Quimby11}. Type I SLSNe (hereafter, \mbox{SLSNe-I}) are classified based on their hydrogen-free spectra, strong \ion{O}{2} absorption lines at early times, and blue continuum \citep{Chomiuk11, Quimby11, PESSTO, Gal-Yam, Inserra13}.  Despite their increasing discovery rate over the past decade thanks to systematic wide-field optical surveys, SLSNe-I still remain a small population, with only about 150 spectroscopically classified events to date \citep{FLEET}; moreover, not all of these events have well observed light curves and spectroscopic time series.

Several possible mechanisms that can power SLSNe-I have been explored in the literature: large radioactive $^{56}$Ni mass \citep{Pastorello10}, circumstellar interaction \citep{Chevalier11}, and a magnetar central engine \citep{Kasen_2010,Woosley10}.  The bulk of the observational evidence points to a magnetar engine as the dominant mechanism.  This includes the diverse light curve behavior \citep{Nicholl17}, the early-time UV spectra \citep{Mazzali16, Nicholl17_16apd, Howell13, Dessart19, Yan17}, the late-time light curve flattening \citep{Nicholl18_1000days, Blanchard2021}, and their nebular spectra \citep{Dessart12,Nicholl19_nebular, Jerkstrand17}.  The preference of SLSNe-I to low metallicity host galaxies (similar to the hosts of the engine-powered long gamma-ray bursts) also supports a magnetar engine energy source \citep{Lunnan14, Leloudas_2015, Orum, Angus19, Perley16, Chen_2017, Schulze18}.

Here, we present detailed optical observations of SN\,2018lfe, a transient first detected by the Zwicky Transient Facility (ZTF; \citealt{Bellm19}) on 2018 December 31 and first reported by the Pan-STARRS Survey for Transients (PSST; \citealt{PSST}) on 2019 January 1. Our follow-up observations of the SN and its host galaxy continue up to a year after discovery and classify the event as a SLSN-I at a redshift of $z=0.3501\pm 0.0004$, with a peak absolute magnitude of $M_r\approx -22.1$. We use this dataset to investigate the properties of SN2018lfe and its host, in the context of magnetar-powered SLSNe, to gain a better understanding of the larger SLSNe population.

The paper is structured as follows. In \S\ref{sec:observations} we present SN\,2018lfe and our multi-band follow-up observations. In \S\ref{sec:spectra} we describe the observed features of the spectra and compare to other SLSNe-I. In \S\ref{sec:modeling} we present our modeling of the light curve using a magnetar engine model, the bolometric light curve, and compare the parameters of SN\,2018lfe to the existing SLSNe-I population. In \S\ref{sec:host} we explore the host galaxy properties. Throughout the paper we assume a flat $\Lambda$CDM cosmology with \mbox{$H_{0} = 69.3$ km s$^{-1}$ Mpc$^{-1}$}, $\Omega_{m} = 0.286$, and $\Omega_{\Lambda} = 0.712$ \citep{hinshaw13}.

\begin{figure}[t!]
	\begin{center}
		\includegraphics[width=\columnwidth]{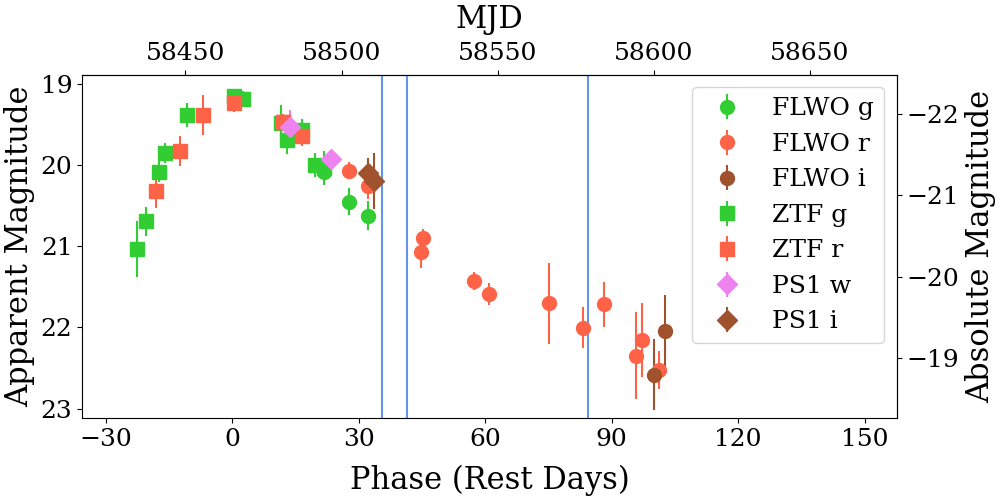}
		\caption{Optical light curve of SN\,2018lfe in the $gwri$ bands using data from FLWO, PSST, and ZTF. The photometry is corrected for Galactic extinction, and time is shown in rest-frame days relative to the time of peak brightness in $r$-band. The blue vertical lines mark the times when the first three spectra were taken (see Section \ref{sec:spectra}), the date of the last spectrum is outside the range of this plot.
			\label{fig:light curve}}	
	\end{center}
\end{figure}

\section{Observations} 
\label{sec:observations}

\subsection{Discovery and Classification}

SN\,2018lfe was first reported by PSST on 2019 January 1 (${\rm MJD}=58484.0$) with a magnitude of $g=19.66$ at coordinates R.A.= ${\rm 09^h33^m29^s.556}$, decl.= $+00^\circ03'08''.39$ (J2000) \citep{Discovery}. We selected SN\,2018lfe for follow-up observations as a probable SLSN-I candidate using our {\it Finding Luminous and Exotic Extragalactic Transients} (FLEET) machine learning pipeline \citep{Gomez_2020}, which determined a probability of SN\,2018lfe being a SLSN-I of $P_{\rm SLSN-I}\approx 0.71$. We obtained a spectrum on 2019 January 31 with the Inamori-Magellan Areal Camera and Spectrograph (IMACS; \citealt{dressler11}) on the Magellan Baade 6.5-m Telescope, and classified it as a SLSN-I \citep{Classification} with aid from template matching to known SN spectra with the Supernovae Identification (SNID, \citealt{SNID}) program. We calculated the redshift \mbox{$z = 0.3501 \, \pm \, 0.0004$} by measuring the center of the H$\beta$ and \mbox{[\ion{O}{3}] $\lambda\lambda$4959,5007} host galaxy emission lines visible in this spectrum in addition to three additional follow-up spectra. 

\subsection{Optical Photometry}
\label{sec:phot}

We obtained images of SN\,2018lfe with the KeplerCam imager on the 1.2-m telescope at Fred Lawrence Whipple Observatory (FLWO) in the $g$, $r$, and $i$ bands, starting on 2019 January 11 (MJD = 58494). SN\,2018lfe was also observed at several earlier epochs by ZTF in the $g$ and $r$ bands, for which we obtained the raw images from the NASA/IPAC Infrared Science Archive\footnote{\url{irsa.ipac.caltech.edu}}. The SN was also observed by PSST in the $w$ and $i$ bands, and we obtained the photometry from the publicly available data on the Open Supernova Catalog\footnote{\label{ref:osc}\url{https://sne.space/}} \citep{guillochon17}.

We processed the FLWO and ZTF images using standard IRAF\footnote{\label{IRAF}IRAF is written and supported by the National Optical Astronomy Observatories, operated by the Association of Universities for Research in Astronomy, Inc. under cooperative agreement with the National Science Foundation.} routines. We modeled the point-spread function from each image using reference stars and subtracted the model from the target to measure the instrumental magnitudes. We combined ZTF images obtained within $\approx 2$ days of each other to improve the signal-to-noise ratio of the measured fluxes. Similarly, for data obtained from FLWO after a phase of 110 days (MJD = 58575), we combined images in the $r$ and $i$ bands that were obtained within $\approx 2$ days of each other. For the PSST photometry, we averaged the reported magnitudes obtained on the same night.

We calibrated all the photometry relative to PS1/3$\pi$ magnitudes and applied a correction for Galactic extinction with \mbox{$E(B-V) = 0.033 \pm 0.001$ mag} and \mbox{$R_V = 3.1 $} \citep{Schlafly11}, using the \cite{Cardelli} Milky Way extinction law. We define phase 0 to be the date of the brightest magnitude in the $r$-band, \mbox{${\rm MJD} = 58465$}, with peak apparent magnitude of $r = 19.2\pm 0.1$ and $g = 19.1\pm 0.1$ after correcting for Galactic extinction. We use this reference point throughout the paper, with all days quoted in the rest frame unless otherwise specified.  The full light curve of SN\,2018lfe is shown in Figure~\ref{fig:light curve}.  The photometry table is provided in the supplementary materials in machine readable format, as well as on the Open Supernova Catalog \citep{guillochon17}.

\subsection{Optical Spectroscopy}
\label{sec:spec}

We obtained four epochs of low-resolution optical spectroscopy at phases of 35, 41, 84, and 256 days. We used the IMACS and the Low Dispersion Survey Spectrograph (LDSS3c, \cite{Stevenson16}) spectrographs on the Magellan 6.5-m telescopes; and the Blue Channel \citep{schmidt89} and Binospec \citep{Fabricant_2019} spectrographs on the MMT 6.5-m telescope. All the spectra were obtained with the slit oriented along the parallactic angle. Details of the spectroscopic observations are provided in Table~\ref{tab:spectroscopy}.

We processed all spectra using standard IRAF routines. A wavelength calibration was applied using a HeNeAr lamp spectrum taken near the time of each science image. Relative flux calibration was applied to each spectrum using a standard star taken on the same night. The spectra were then calibrated to absolute flux scaling using the available $gri$ photometry. Lastly, the spectra were corrected for Galactic extinction and transformed to the rest frame of SN\,2018lfe.

The resulting spectra are shown in Figure~\ref{fig:spectra}, with prominent emission features from the host galaxy, absorbtion features of the SN, and telluric absorption features marked. The spectra are also available on the Weizmann Interactive Supernova Data Repository (WISeREP\footnote{\label{ref:wiserep}\url{https://wiserep.weizmann.ac.il}}; \citealt{WISeREP}).

\startlongtable
\begin{deluxetable*}{lcccccccc}
	\tabletypesize{\footnotesize}	\tablecaption{Optical Spectroscopy of SN\,2018lfe \label{tab:spectroscopy}}
	\tablewidth{0pt}
	\tablehead{
		\colhead{UT Date} & 
		\colhead{MJD} & 
		\colhead{Phase\tablenotemark{a}} & 
		\colhead{Exp.~Time} & 
		\colhead{Airmass} & 
		\colhead{Wavelength} & 
		\colhead{Telescope + Instrument} & 
		\colhead{Grating} & 
		\colhead{Slit Width} \\
		\colhead{} & 			  
		\colhead{} & 
		\colhead{(d)}   & 
		\colhead{(s)} &			  
		\colhead{} &		
		\colhead{(\AA)}	&		
		\colhead{} & 
		\colhead{(lines/mm)} &		 
		\colhead{($''$)}						 
	}
	\startdata
    2019 Jan 30  & 58513  & 35 & 1200    & 1.17    & 4500--9400  & Baade + IMACS      & 300     & 0.9   \\
    2019 Feb 7 & 58521  & 41 & 1200    & 1.50    & 4500--9500  & Clay + LDSS & 300     & 1      \\
    2019 Apr 6  & 58579  & 84  &  1800 & 1.16 & 4500--9400  & Baade + IMACS        & 1600 & 1 \\
    2019 Nov 24 & 58811  & 256 & 900  & 1.22  & 3800--9200  & MMT + Binospec & 300     & 0.9      \\
   	\enddata
	\tablenotetext{a}{Rest-frame days since peak r-band luminosity.}
\end{deluxetable*}

\section{Spectral Features and Comparisons}
\label{sec:spectra} 

The earliest spectrum of SN\,2018lfe (35 days) exhibits a blue continuum and broad absorption features. We detect a broad feature at the location of \ion{Mg}{1]} $\lambda$4500 and measure its width with a Gaussian profile, leading to velocity widths of $11190\pm 660$, $ 11770\pm 600$, and $10180\pm 470$ km s$^{-1}$ for phases of 35, 41, and 84 days, respectively. There is thus no evidence for a significant velocity evolution. The three early spectra also show \ion{Si}{2} $\lambda$6355  regions with an average velocity center of $\approx 11700$ km s$^{-1}$. This is consistent with the ejecta velocity obtained from fitting the light curve with a magnetar model (see \S\ref{sec:modeling}). As is common for SLSNe-I, the continuum becomes redder at later phases. No obvious SN features are detected in the final spectrum at 256 days, which shows only narrow emission lines from the host galaxy.

\begin{figure*}
	\begin{center}
		\includegraphics[width=\textwidth]{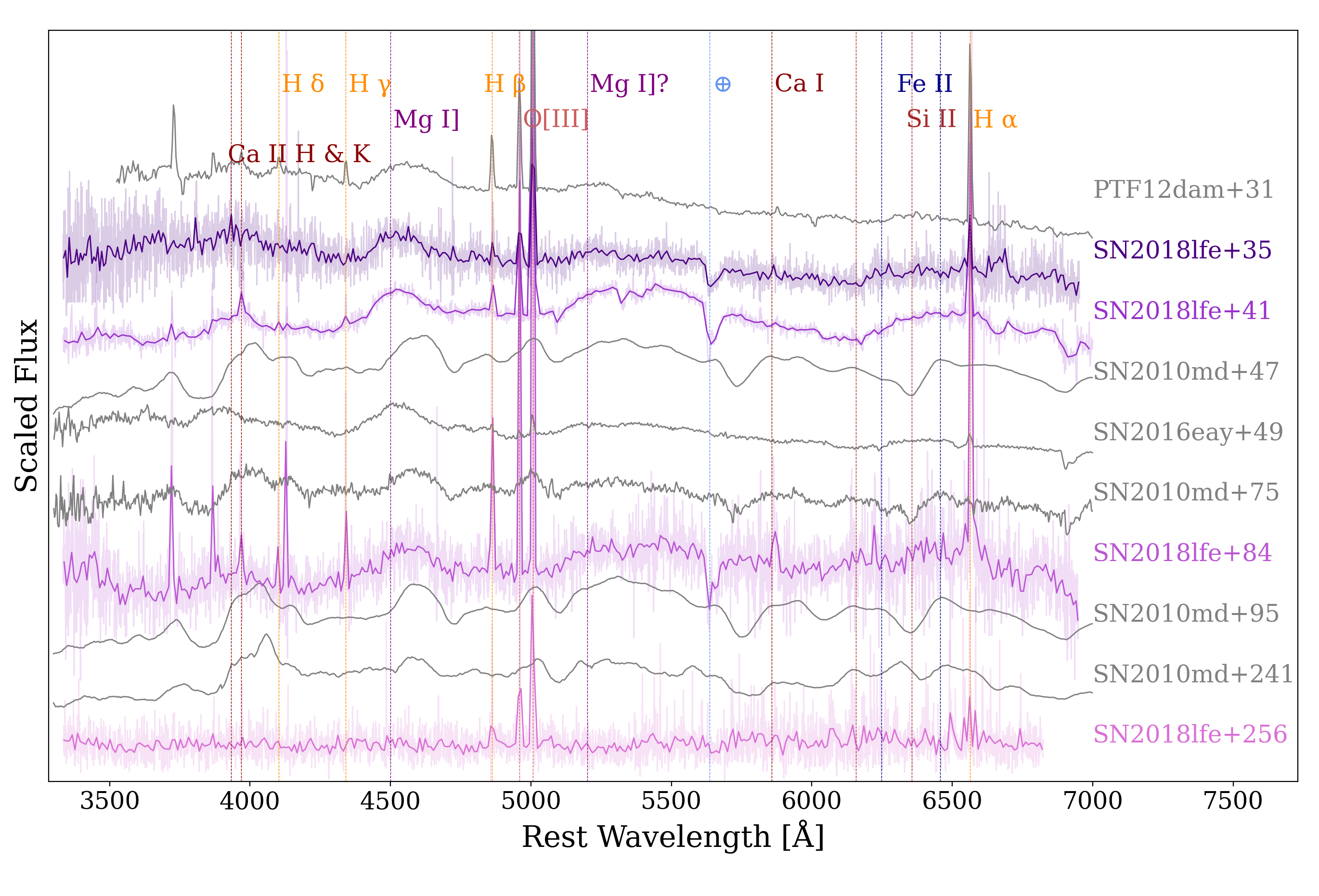}
		\caption{Optical spectra of SN\,2018lfe spanning a rest frame phase of 35 to 256 days, with the last spectrum being dominated by host galaxy emission. The spectra are corrected for Galactic extinction and shifted to the rest-frame of SN\,2018lfe using $z=0.3501$. We bin the spectra for clarity and show the unbinned spectra as shaded regions. The labeled vertical lines mark spectral features from the SN, narrow lines from the host galaxy, and telluric absorption. The different colors in the vertical lines represent different elements. We compare the evolution of SN\,2018lfe to three other SLSNe-I: SN\,2010md (PTF10hgi; \citealt{Inserra13}), SN\,2016eay (Gaia16apd; \citealt{Nicholl_2017}), and PTF12dam \citep{Nicholl13}, with their spectra colored gray. \label{fig:spectra}}
	\end{center}
\end{figure*}

We compare the spectra of SN\,2018lfe to those of three other SLSNe-I in Figure~\ref{fig:spectra}. The objects SN\,2010md (=PTF10hgi) \citep{Inserra13}, SN\,2016eay \\ (=Gaia16apd) \citep{Nicholl_2017}, and PTF12dam \citep{Nicholl14} were chosen for having similar values of magnetar engine and ejecta parameters\footnote{The values of these parameters are from our previous work in \citet{Nicholl17}.} as SN\,2018lfe, and for their abundance of spectral data, specifically at phases that match those of SN\,2018lfe. 

The emission regions of \ion{Mg}{1]} $\lambda$4571, \ion{Ca}{2} H $\lambda$3933,  \ion{Ca}{2} K $\lambda$3968 in the early spectrum of SN\,2018lfe at phase of $35-41$ days also match with the early phase data of SN\,2010md and PTF12dam. As demonstrated in Figure \ref{fig:spectra}, the spectroscopic evolution of SN\,2018lfe follows that of SN\,2010md. In particular, both objects display prominent \ion{Mg}{1}] and \ion{Ca}{2} emissions features that are similar in shape at similar phases, namely $41-47$, $75-95$, and $241-256$ days from explosion. However, SN2010md shows an unusually prominent absorption line near 5700 \AA, unlike SN\,2018lfe and other typical SLSNe-I.

\section{Light Curve Modeling} 
\label{sec:modeling} 

\begin{deluxetable*}{ccccc}
	\tablecaption{ Best-fit Parameters for the Magnetar Models \label{tab:magnetar}}
	\tablehead{\colhead{Parameter} & \colhead{Definition} & \colhead{Prior}  & \colhead{Full Light Curve} & \colhead{Units}}
	\startdata
	$\log(M_{\text{ej}})$ & Ejecta Mass & $[-1, 2]$ & $ 0.58^{+0.11}_{-0.11}$ & M$_\odot$ \\
	$B_{\perp}$ & Magnetar Magnetic Field Strength & $[0.1, 10]$ & $ 2.8^{+2.1}_{-1.0}$ &  $10^{14} $G \\
	$P_{\text{spin}}$ & Magnetar Spin & $[1, 10]$ & $ 2.89^{+0.47}_{-0.58}$ &  \\
	$M_{\text{NS}}$ & Mass of Neutron Star & $[0.1, 2.2]$ & $ 1.87^{+0.24}_{-0.41}$ & M$_\odot$ \\
	$v_{\text{ej}} $ & Ejecta Velocity & $[5.0e3,2.0e4]$ & $ 10062^{+867}_{-949}$ & km s$^{-1}$ \\
	$E_{k}^{\dagger}$ & Ejecta Kinetic Energy & & $ 3.9^{+1.3}_{-1.2}$ & $10^{51}$ erg s$^{-1}$ \\
	$t_{\text{exp}} $ & Explosion Time Relative To First Data Point & $[-500, 0]$& $ -5.4^{+0.86}_{-1.0}       $ & days                   \\
    $\log(n_{H,\rm host}) $ & Column density in the host galaxy & $[16, 23]$& $ 18.4^{+1.7}_{-1.7}       $ & cm$^{-2}$                   \\
	$T_{\text{min}}$ & Photospheric Temperature & $[3.0e3, 1.0e4]$  & $ 7683^{+1532}_{-1731}$ &   K                      \\
	$\kappa $ & Optical opacity  & $[0.05, 0.2]$       & $ 0.16^{+0.02}_{-0.04}  $ &  cm$^2$g$^{-1}$          \\
	$\log(\kappa_{\gamma}) $ & Gamma-ray opacity & $[-1, 4]$ & $ 1.6 \pm 1.6 $ & cm$^2$g$^{-1}$ \\
	\enddata
	\tablecomments{Best model parameters, definitions, prior ranges, and 1$\sigma$ error bars for the realizations shown in Figure~\ref{fig:lc_mosfit}.}
	\tablenotetext{\dagger}{These parameters were not fit for, but were calculated using all the posterior distribution samples of the fitted parameters.}
	\end{deluxetable*}
	
We model the multi-band light curve with a magnetar spin-down model implemented in the Modular Open-Source Fitter for Transients ({\tt MOSFiT}) Python package, a Markov chain Monte Carlo (MCMC) code that can fit the light curves of transients using a variety of power sources \citep{guillochon18}. We use the {\tt emcee} \citep{foreman13} implementation of MCMC to run each sampler. Since {\tt MOSFiT} implements a simplified analytical one-zone model, the uncertainties presented here represent only the statistical errors on the fit. The priors used in the model are listed in Table~\ref{tab:magnetar}, following our previous work in \citet{Nicholl_2017}.

\begin{figure}
	\begin{center}
		\includegraphics[width=0.95\columnwidth]{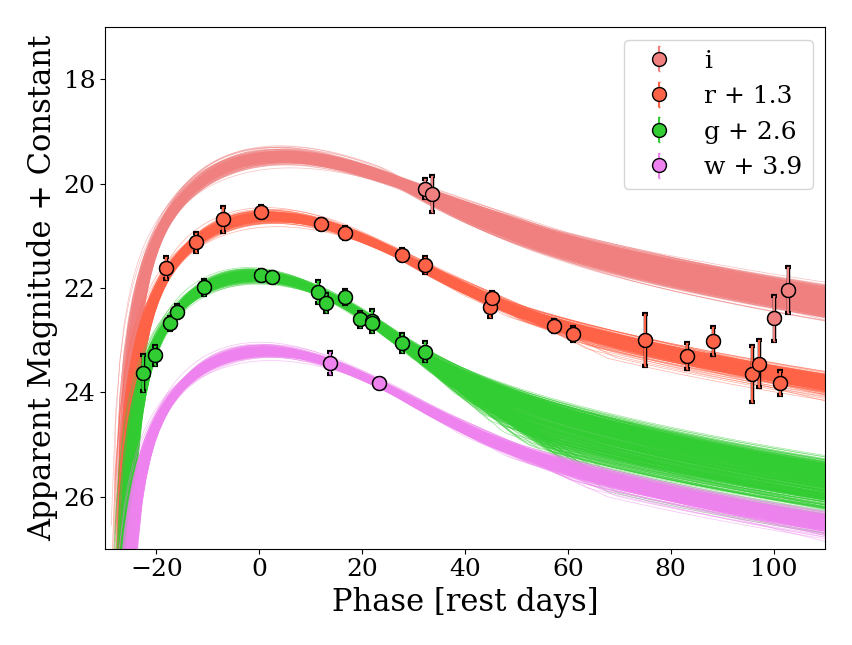}		
		\caption{{\it Top Panel:} Light curve of SN\,2018lfe with a magnetar engine model. Each line is a sample realization of the most likely models generated from {\tt MOSFiT}. 
		}
		\label{fig:lc_mosfit} 
	\end{center}
\end{figure}

\begin{figure}[hb]
	\begin{center}
		\includegraphics[width=0.48\textwidth]{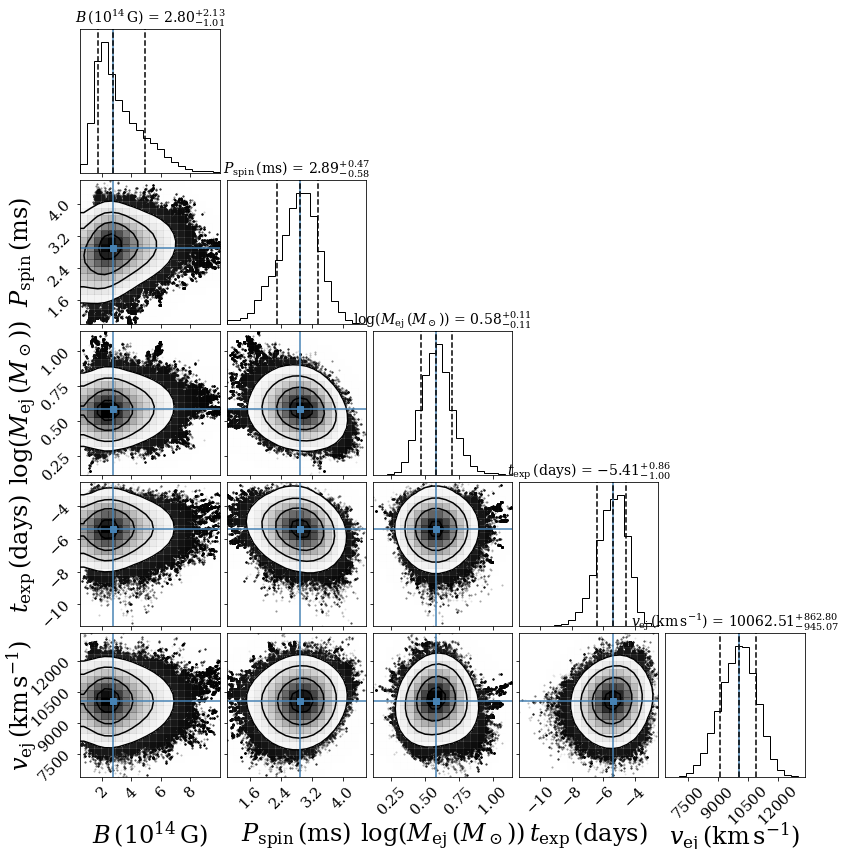}
		\caption{\label{fig:corner} Posterior distributions and correlations of the magnetar model parameters. The median and $1\sigma$ values are marked and labeled. A summary of all the parameter values is presented in Table~\ref{tab:magnetar}.}
	\end{center}
\end{figure}

The magnetar model fits are shown in Figure~\ref{fig:lc_mosfit}, and the parameter posterior distributions are shown in Figure~\ref{fig:corner} and listed in Table~\ref{tab:magnetar}. The model provides an excellent fit to the data, with the following primary physical parameters: an initial spin period of \mbox{$P\approx 2.9$ ms}, a magnetic field of \mbox{$B_{\perp}\approx 2.8\times 10^{14}$ G}, an ejecta mass of \mbox{$M_{\rm ej}\approx 3.8$ M$_\odot$}, and an ejecta velocity of $v_{\rm ej}\approx 10^4$ km s$^{-1}$ (in good agreement with the velocity inferred from the spectra in \S\ref{sec:spectra}). We calculate a total kinetic energy of \mbox{$E_K\approx 4\times 10^{51}$ erg}. The mass of the progenitor star is estimated to be $M_{\rm NS}+M_{\rm ej}\approx 5.7$ M$_{\odot}$, which is consistent with the range of $3.6-40$ M$_{\odot}$ inferred from the SLSNe-I sample of \citet{Blanchard20}.

\begin{figure}
	\begin{center}
		\includegraphics[width=\columnwidth]{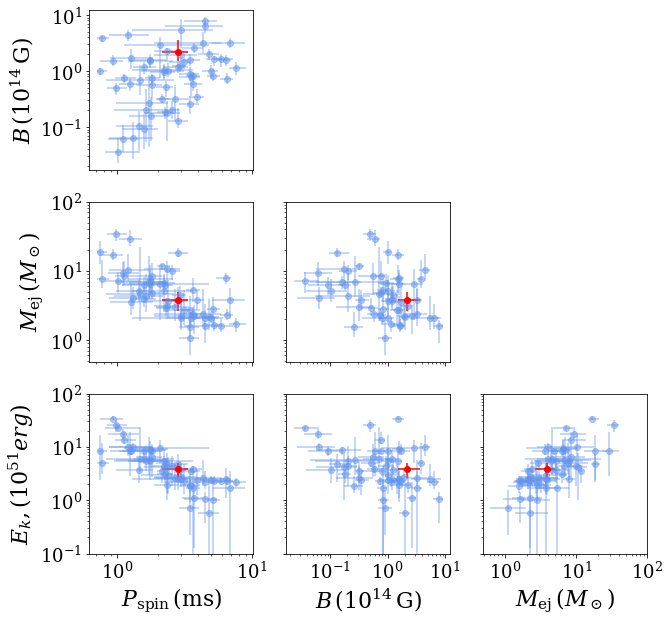}
		\caption{\label{fig:mosfit_params} Median values and 1$\sigma$ errors of key parameters (P, $B_{\perp}$, $M_{\rm ej}$, $E_K$ ) for all SLSNe in \citealt{Blanchard20, Villar18} and \citealt{Nicholl15} (labeled blue) compared with SN\,2018lfe (labeled orange). SN\,2018lfe lies consistent with the magnetar model.}
	\end{center}
\end{figure}

We compare the resulting parameters for SN\,2018lfe to the distribution of $62$ SLSNe from \citet{Nicholl15, Blanchard20, Villar18} in Figure~\ref{fig:mosfit_params}. As seen in the Figure, all of the parameters for SN\,2018lfe fall within the range covered by previous SLSNe. However, we do find that the magnetic field value inferred for SN\,2018lfe is in the top $16^{\rm th}$ percentile of the SLSN distribution. The combination of the strong magnetic field, fast spin ($P_{\rm spin} < 3$ ms), and high ejecta velocity is most likely responsible for  significant heating of the ejecta, which explains the high optical luminosity of SN\,2018lfe.

\begin{figure}
	\begin{center}	\includegraphics[width=0.97\columnwidth]{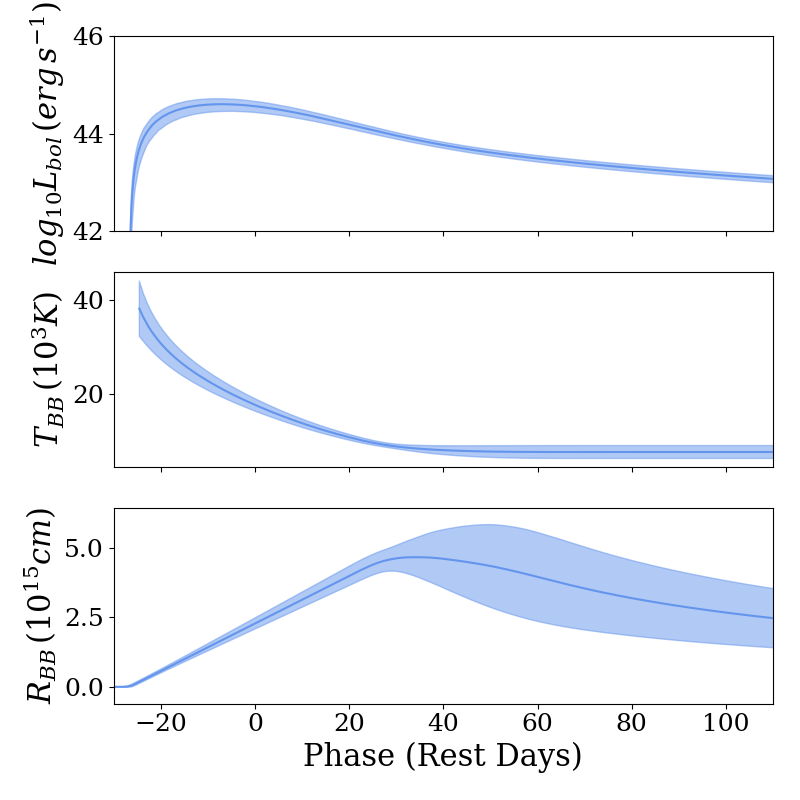}
		\caption{Bolometric light curve, blackbody temperature evolution, and photospheric radius evolution obtained from the \texttt{MOSFiT} models. The plots show the mean model (blue line) and the $1\sigma$ regions of 500 walkers used in the MCMC process.
		}
		\label{fig:bolometric_mosfit} 
	\end{center}
\end{figure}

We further use the \texttt{MOSFiT} model to construct a bolometric light curve, photospheric radius evolution, and temperature evolution of SN\,2018lfe; see Figure~\ref{fig:bolometric_mosfit}.  The shaded regions in the Figure represent the $1\sigma$ values across all 500 walkers. We find a peak bolometric luminosity of \mbox{$(4.0\pm 1.1) \times 10^{44}$ erg s$^{-1}$}. The integrated radiated energy over the observed portion of the light curve (phase 0 to 103 days) is $E_{\rm rad}\approx 1.2\times 10^{51}$ erg, consistent with the known population of SLSNe, despite its bright peak luminosity \citep{Lunnan18, Nicholl20_nature}. The photospheric radius exhibits an increase up to a maximum of $4.6^{+1.1}_{-0.5}\times 10^{15}$ cm at a phase of $\approx 30-50$ d, with a gradual decline thereafter. Finally, the photospheric temperature does shows a steady decline from an inferred peak of $\approx 3.5\times 10^4$ K down to $\approx 10^4$ K at a phase of $\gtrsim 30$ days and suggest a temperature $<10^4$ K at 40 days. Although due to the lack of data in the UV range, there is large uncertainty at early times when the spectral energy distribution peaks in this wavelength range.

\section{Host Galaxy}
\label{sec:host} 

No galaxy is detected at the position of SN\,2018lfe in the PS1/$3\pi$ catalog down to the nominal $5\sigma$ limiting magnitudes of $g\approx 23.3$, $r\approx 23.2$, $i\approx 23.1$, and $z\approx 22.3$ \citep{Chambers2019}. We obtained deeper images on 2020 November 15 using Binospec and detected an extended source with a half-light radius of $\approx 0.5''$, measured using \texttt{sep} \citep{sep}. From extrapolating the observed light curve (see Section \ref{sec:phot}), we expect the apparent magnitude of the SLSN to be dimmer than $\approx 25$ mag in the r-band, therefore the host should be the only source present. We extracted the host magnitudes by performing aperture photometry using the {\tt photutils} package \citep{larry_bradley_2020_4044744} with a circular aperture of radius $2''$ across all filters. At the measured redshift of $z=0.3501$, the observed $r$-band is closest to rest-frame $B$-band. The apparent host magnitude of $r\approx 23.52$ mag thus corresponds to a $L_B\approx 0.029$ L$^*$, determined using the DEEP2 luminosity function at $z=0.3$ \citep{Willmer_2006}. The host galaxy properties are summarized in Table~\ref{tab:host}.

We measured the strength of the galaxy emission lines by fitting narrow Gaussian profiles (FWHM$\approx 40$\AA) to the three spectra at phases of 35, 41, and 84 days. The flux ratio of \mbox{$L_{\rm H\alpha}/L_{\rm H\beta}=2.75\pm 0.37$} is consistent with the expectation for case B recombination ($L_{\rm H\alpha}/L_{\rm H\beta}=2.86$), suggesting no significant host galaxy extinction (in agreement with the lack of extinction inferred from modeling of the SN light curves). The H$\alpha$ line luminosity is \mbox{$L_{\rm H\alpha}= (9.8\pm 0.3)\times 10^{43}$ erg s$^{-1}$}.  This corresponds to a star formation rate (SFR) of $0.77\pm 0.02$ M$_\odot$ yr$^{-1}$ \citep{kennicutt98} 

To calculate the metallicity we use the $R_{23}$ method \citep{kobulnicky99} and adopt the lower branch solution due to the lack of an \ion{N}{2} detection. We determine a value of \mbox{$12 + \log($O/H$)=8.18 \pm 0.12$}, or \mbox{$Z = 0.31\pm 0.05$ Z$_\odot$} \citep{asplund09}. Coincidentally, this is also where the two values from the $R_{23}$ method converge, and therefore our metallicity is not affected by the choice of the lower or upper branch solution. The low metallicity is consistent with the consensus that SLSNe preferentially occur in galaxies with sub-solar metallicity  \citep{Lunnan14,Chen_2017,Perley16,Schulze_2017}. 

\begin{deluxetable}{ccc}
	\tablecaption{Host Galaxy Properties \label{tab:host}}
	\tablehead{\colhead{} & \colhead{Value (mag)} & \colhead{Units}}
	\startdata
	$g$ &  $24.07 \pm 0.20$ & mag \\
    $r$ &  $23.52 \pm 0.24$ & mag \\
    $i$ &  $24.01 \pm 0.19$ & mag \\
    $z$ &  $24.36 \pm 0.19$ & mag \\
    $M_g$ &  $-17.30 \pm 0.2$ & mag \\
    $M_r$ &  $-17.85 \pm 0.3$ & mag \\
    $M_i$ &  $-17.36 \pm 0.4$ & mag \\
    $M_z$ &  $-17.01 \pm 0.3$ & mag \\
    $L_B$ & $0.029 \pm 0.01$ & $L^*$ \\
    $12 + \log($O/H$)$ &  $8.18\pm0.12$ & \\
    $Z$ &  $0.31 \pm 0.05$ & $Z_\odot$  \\
    ${\rm SFR}$ & $0.77 \pm 0.02$  & M$_\odot$ yr$^{-1}$  \\
	\enddata
	\tablecomments{Apparent magnitudes, not corrected for extinction, obtained from observations using the Binospec \citep{fabricant03} imaging camera on the MMT on November 15, 2020. Absolute magnitudes are corrected for extinction. The star formation rate and the metallicities (in terms of $12 + \log($O/H$)$ and Z$_\odot$) are estimated using the flux immensities of the host galaxy emission lines. $L_B$ is the host luminosities relative to $L^*$ in the $B$ band \citep{Willmer_2006}.}, due to the fact that at $z=0.35$, the observed $r$-band is closest to rest-frame B-band, see Section~\ref{sec:host}.
\end{deluxetable}

\section{Conclusions} 
\label{sec:conclusion}

We presented the classification and detailed optical follow-up observations of the Type-I SLSN 2018lfe and its host galaxy at $z = 0.3501$. We model the light curve using a spin down magnetar model and find that SN\,2018lfe has a best fit ejecta mass of $\approx 3.8$ M$_\odot$, a magnetar spin period of $\approx 2.9$ ms and a magnetic field strength of $\approx 2.8\times 10^{14}$ G, consistent with the population of known SLSNe. We determine the host galaxy properties through spectral analysis and follow up photometric observations and find a low metallicity of \mbox{$12 + \log($O/H$)=8.18 \pm 0.12$}, typical for SLSNe-I hosts. Therefore, SN\,2018lfe is a typical SLSN-I in terms of its explosion and host galaxy properties and contributes to a small population of SLSNe-I with a well-defined explosion time and peak absolute magnitude. 

\acknowledgments
We thank Y.~Beletsky for carrying out some of the Magellan observations. The Berger Time-Domain Group at Harvard is supported in part by NSF under grant AST-1714498 and by NASA under grant NNX15AE50G. M.~Nicholl is supported by a Royal Astronomical Society Research Fellowship and by the European Research Council (ERC) under the European Union’s Horizon 2020 research and innovation programme (grant agreement No.~948381).  Operation of the Pan-STARRS1 telescope is supported by the National Aeronautics and Space Administration under grant No. NNX12AR65G and grant No. NNX14AM74G issued through the NEO Observation Program. This paper includes data gathered with the 6.5 meter Magellan Telescopes located at Las Campanas Observatory, Chile. Observations reported here were obtained at the MMT Observatory, a joint facility of the University of Arizona and the Smithsonian Institution. This research has made use of NASA’s Astrophysics Data System. This research has made use of the SIMBAD database, operated at CDS, Strasbourg, France.

\software{Astropy\citep{astropy18}, MOSFiT\citep{guillochon18}, PyRAF\citep{science12}, SAOImage DS9 \citep{Smithsonian00}, emcee\citep{foreman13}, corner \citep{foreman16}, Matplotlib\citep{hunter07}, SciPy\citep{Walt11}, NumPy\citep{Oliphant07}, SEP\citep{sep} extinction\citep{Barbary16}}, FLEET\citep{FLEET}).


\bibliography{main.bib}

\end{document}